\documentclass{article}

\usepackage{arxiv}
\usepackage[utf8]{inputenc}
\usepackage[T1]{fontenc}
\usepackage{hyperref}
\usepackage{url}
\usepackage{booktabs}
\usepackage{amsfonts}
\usepackage{amsmath}
\usepackage{graphicx}
\usepackage{algorithm}
\usepackage{algorithmic}
\usepackage{subcaption}
\usepackage{xcolor}
\usepackage{siunitx}

\title{City-Scale Visibility Graph Analysis via GPU-Accelerated HyperBall}

\author{
  Alex Hodge \\
  Independent Researcher, London, UK \\
  \texttt{alex.hodge@gmail.com} \\
  \And
  Dr.\ Melissa Barrientos Tri\~{n}anes \\
  School of Geography, University of Leeds, UK \\
  \texttt{M.R.BarrientosTrinanes@leeds.ac.uk}
}

\begin{document}
\maketitle

\begin{abstract}
Visibility Graph Analysis (VGA) is a key space syntax method for understanding
how spatial configuration shapes human movement, but its reliance on
all-pairs BFS computation limits practical application to small study areas.
We present a system that combines three techniques to scale VGA to city-scale
problems: (i)~delta-compressed CSR storage using LEB128 varint encoding, which
achieves ${\sim}4{\times}$ compression and enables memory-mapped graphs
exceeding available RAM; (ii)~HyperBall, a probabilistic distance estimator
based on HyperLogLog counter propagation, applied here for the first time to
visibility graphs, reducing BFS complexity from $O(N \cdot |E|)$ to
$O(D \cdot |E| \cdot 2^p)$; and (iii)~GPU-accelerated CUDA kernels with a fused
decode-union kernel that streams the compressed graph via PCIe and performs
LEB128 decoding entirely in shared memory.
HyperBall's iteration count equals the topological depth limit,
so the radius-$n$ analysis that practitioners already use as
standard~\cite{hillier1996space} translates directly into proportional
speedup---unlike depthmapX, whose BFS time is invariant to depth setting
due to the small diameter of visibility graphs.
Using depthmapX's own visibility algorithm (sparkSieve2) to ensure identical
edge sets, our tool achieves a $239{\times}$ end-to-end speedup at
\num{42705}~cells and scales to \num{236000}~cells (\num{4.8}~billion edges)
in \SI{137}{\second}---problem sizes far beyond depthmapX's practical limit.
At $p = 10$, Visual Mean Depth achieves Pearson $r = 0.999$ with
\SI{1.7}{\percent} median relative error across 20 matched configurations.
\end{abstract}


\section{Introduction}
\label{sec:introduction}

Space syntax~\cite{hillier1984social} studies how the geometric configuration
of space shapes human behaviour: movement patterns, social co-presence, and
land-use distribution.
Visibility Graph Analysis (VGA)~\cite{turner2001isovists} extends space syntax
from street networks to continuous open space by placing a regular grid of
points over the study area, connecting mutually visible pairs, and computing
graph-theoretic metrics on the resulting structure.
Metrics such as visual integration, mean depth, and clustering coefficient
reveal spatial properties---visual accessibility, topological remoteness,
convex-versus-linear character---that are invisible to purely geometric
analysis.

The computational cost of VGA grows rapidly with problem size.
The key bottleneck is all-pairs shortest-path computation: metrics like mean
depth require BFS from every node, costing $O(N \cdot |E|)$ for the main
connected component.
At fine grid resolution (\SIrange{1}{5}{\metre}) over city-scale areas
($> \SI{1}{\kilo\metre\squared}$), graphs reach tens of thousands of nodes
with hundreds of millions of edges.
depthmapX~\cite{varoudis2012depthmapx}, the standard tool, is
single-threaded and in-memory; in our benchmarks it scales as $O(N^2)$ and
timed out at \SI{30}{\minute} beyond \num{42705} cells.

We present a system that overcomes these limits through three contributions:
\begin{enumerate}
\item \textbf{Delta-compressed CSR storage.}
Neighbour indices are delta-encoded as LEB128 varints, achieving
${\sim}4{\times}$ compression.
Memory-mapping allows graphs exceeding RAM to be processed.

\item \textbf{HyperBall for VGA with depth-proportional speedup.}
We apply HyperBall~\cite{boldi2013hyperball}---a probabilistic distance
estimator based on HyperLogLog counter propagation---to visibility graphs for
the first time, reducing all-pairs BFS from $O(N \cdot |E|)$ to
$O(D \cdot |E| \cdot 2^p)$ where $D$ is the graph diameter.
Unlike depthmapX's BFS, which runs in constant time regardless of depth
setting due to the small diameter of visibility graphs, HyperBall converges
in exactly $\min(d, D)$ iterations.
At depth limit~3---the standard local measure in VGA
practice~\cite{turner2004depthmap,koutsolampros2019dissecting}---this yields a
$352{\times}$ BFS-phase speedup.

\item \textbf{GPU-accelerated CUDA pipeline.}
A fused decode-union kernel streams the compressed graph to the GPU and
performs LEB128 decoding in shared memory.
\end{enumerate}

To ensure a fair comparison, we use depthmapX's own visibility algorithm
(sparkSieve2 angular sweep), ported to Rust with data-parallel execution via
Rayon, producing identical edge sets and isolating accuracy differences to the
HLL approximation alone.
Across 20 matched configurations, Visual Mean Depth achieves Pearson
$r = 0.999$ at HLL precision $p = 10$ with \SI{1.7}{\percent} median relative
error, while our tool delivers a $239{\times}$ end-to-end speedup at
\num{42705} cells and scales to \num{236000} cells (\num{4.8} billion edges)
in \SI{137}{\second}.
\section{Background and Related Work}
\label{sec:background}

\subsection{Space Syntax and VGA}
\label{sec:background:spacesyntax}

Space syntax~\cite{hillier1984social} is a family of theories and analytical
methods that relate the geometric configuration of space to patterns of human
movement and co-presence.
The core insight is that the \emph{topology} of the street network---how each
segment connects to every other---predicts where people walk, where retail
activity concentrates, and which spaces feel lively or deserted, often more
strongly than deliberate planning interventions.

Visibility Graph Analysis (VGA), introduced by Turner et al.~\cite{turner2001isovists}
and building on Benedikt's isovist framework~\cite{benedikt1979isovists},
extends space syntax from street networks to continuous open space.
A regular grid of points is placed over the study area; points inside buildings
are removed; and any two remaining points that share an unobstructed line of
sight within a radius $r$ are connected by an edge in the \emph{visibility
graph}.
Graph-theoretic metrics are then computed on this structure to characterise each
location's visual integration within the whole.
The thirteen standard metrics span local and global properties:
\emph{connectivity} (immediate degree); \emph{visual mean depth} (mean
shortest-path distance); three \emph{integration} variants (Hillier-Hanson
[HH]~\cite{hillier1984social}, Teklenburg [Tekl]~\cite{teklenburg1993space},
and P-value~\cite{turner2004depthmap}); \emph{control} and \emph{controllability}
(local visual dominance); \emph{clustering coefficient}~\cite{watts1998collective};
visual \emph{entropy} and \emph{relativised entropy}; and \emph{point first} and
\emph{second moments}~\cite{turner2001isovists}.
Visually integrated locations---those with low mean depth and high
integration---attract pedestrian movement and anchor social activity;
clustering coefficient distinguishes convex plaza-like spaces from linear
corridor-like spaces~\cite{turner2001isovists,koutsolampros2019dissecting}.
VGA has been applied to urban public space, building interiors, and, as in the
present work, city-scale open-space analysis~\cite{zumelzu2019valdivia}.

\subsection{Computational Challenges in VGA}
\label{sec:background:computation}

VGA poses three compounding computational challenges at city scale.

\paragraph{Graph density.}
The mean degree of a visibility graph node grows as $\pi(r/s)^2$ times the
open-space occupancy fraction, where $r$ is the visibility radius and $s$ is
the grid spacing.
At \SI{2}{\metre} spacing and \SI{300}{\metre} radius the expected mean degree
is approximately \num{20800}, giving roughly \num{60} billion directed edges for
\num{2.9} million nodes.
In standard 32-bit CSR format this requires approximately \SI{240}{\giga\byte}
for the index array alone.

\paragraph{BFS metric cost.}
Mean depth and all derived integration metrics require all-pairs shortest-path
distances.
Exact BFS from every node costs $O(N \cdot |E|)$; at $N = 2.9 \times 10^6$
with billions of edges, exact computation is infeasible on commodity hardware.
The standard tool depthmapX~\cite{varoudis2012depthmapx} is a
single-threaded, in-memory implementation whose practical limit is
approximately \num{50000}--\num{100000} nodes.

\paragraph{Existing approximations and their limitations.}
Tiling---processing overlapping spatial sub-regions independently---was an
early strategy for reducing memory demand, but BFS metrics are truncated at
tile boundaries, introducing severe artefacts.
Testing on a \SI{950}{\metre} study area with \SI{200}{\metre} tiles showed
\SI{84}{\percent} error in Visual Node Count and \SI{59}{\percent} error in
Mean Depth.
The errors are structural, not implementation-specific, and arise because
boundary truncation shortens apparent shortest paths and artificially inflates
integration values.
Viraph~\cite{aminibehbahani2017viraph} avoids full BFS by decomposing the
floor plan into convex sub-areas and computing interspatial depth via weighted Dijkstra on the resulting
convex-region graph, enabling faster analysis for small study areas.
Landmark BFS~\cite{eppstein2004fast,potamias2009fast}---running exact BFS from
$K \approx \sqrt{N}$ spatially stratified source nodes and averaging the
resulting distances---provides artefact-free approximations with
$O(1/\sqrt{K})$ convergence and $3$--$5\%$ relative error, but requires
$K \cdot |E|$ edge traversals: at $K = 1709$ and \num{60} billion edges, this
takes more than \SI{24}{\hour} on the hardware described in
Section~\ref{sec:evaluation:setup}.

\subsection{HyperLogLog and HyperBall}
\label{sec:background:hyperball}

HyperLogLog (HLL)~\cite{flajolet2007hyperloglog} is a probabilistic
cardinality estimator that represents a set as $m = 2^p$ registers, each
storing the maximum leading-zero count seen among hashed elements.
It estimates set cardinality from the harmonic mean of $2^{-\text{register}}$
values with standard error $1.04 / \sqrt{m}$; at $p = 8$ ($m = 256$) the
standard error on raw cardinality is approximately \SI{6.5}{\percent}.
The union of two HLL sketches is computed in $O(m)$ time by taking the
element-wise maximum of their register arrays---a property that enables
efficient propagation over graph edges.

HyperBall~\cite{boldi2011hyperanf,boldi2013hyperball} exploits this property to
estimate the \emph{neighbourhood function} $|B(v, t)|$---the number of nodes
within $t$ hops of $v$---for every node simultaneously.
In each iteration $t$, each node's HLL counter is unioned with the counters of
all its neighbours; after $t$ iterations, counter $v$ encodes an estimate of
$|B(v, t)|$.
The key identity relating the neighbourhood function to mean depth is
\begin{equation}
  \sum_{u \neq v} d(v, u)
    = \sum_{t=1}^{D} t \cdot \bigl(|B(v,t)| - |B(v,t-1)|\bigr),
  \label{eq:hbidentity}
\end{equation}
where $D$ is the diameter.
This allows mean depth to be computed from the sequence of cardinality estimates
accumulated across iterations, without storing any per-node distance array.
The algorithm converges in $O(D)$ iterations, each requiring one scan over all
edges to propagate $m = 2^p$ registers per union, for total time
$O(D \cdot |E| \cdot 2^p)$.
HyperBall has been applied to web and social graphs at scales of billions of
nodes~\cite{boldi2013hyperball}; to our knowledge, this work is the first
application to visibility graphs.

\section{Method}
\label{sec:method}

\subsection{Visibility Graph Construction}
\label{sec:method:visibility}

The construction pipeline takes building footprints and a study area boundary as
input and produces a delta-compressed CSR graph.
Building edges are first rasterised into a per-cell CSR structure at the grid
spacing resolution; grid points are sampled at spacing $s$ and filtered to
exclude cells inside building footprints, yielding the set of visibility
nodes~$V$.

\paragraph{SparkSieve angular sweep.}
Visibility is determined using the sparkSieve2 algorithm from
depthmapX~\cite{varoudis2012depthmapx}, ported to Rust.
For each source cell, eight octants expand outward ring-by-ring, maintaining a
list of angular gaps in $[0, 1]$ tan-space.
At each depth ring, obstacle line segments crossing the ring are projected into
tan-space and subtracted from the gap list; grid cells that fall within
remaining gaps are visible and added to the neighbour list.
When all gaps close, the octant terminates.
Work is proportional to the number of visible cells, not the search area,
making the algorithm efficient in obstacle-rich environments.

Using depthmapX's own visibility algorithm is a deliberate design choice:
it produces identical edge sets, isolating all accuracy and performance
differences to the HyperBall BFS approximation alone.
Source nodes are processed in parallel via Rayon~\cite{rayon2024}; neighbour
lists are delta-encoded and appended to the compressed CSR store in batched
writes.
Connected components are computed incrementally during construction via
Union-Find (path halving, union by rank), requiring no post-hoc graph
traversal.

\subsection{Delta-Compressed CSR Representation}
\label{sec:method:csr}

Standard CSR stores neighbour indices as a flat array of 32-bit integers.
At \num{60} billion edges this requires approximately \SI{240}{\giga\byte} for
the index array alone---far exceeding available RAM and precluding in-memory
graph operations.
The delta-compressed CSR format solves both the storage problem and, crucially,
enables GPU-accelerated HyperBall by keeping the on-wire graph size within the
PCIe streaming budget (Section~\ref{sec:method:gpu}).

The key observation is that visibility graph neighbour lists, when sorted by
node index, have small successive differences.
Nodes are numbered in raster scan order, so neighbours within the same row of
the grid differ in index by 1 or 2; neighbours in adjacent rows differ by the
grid width.
At \SI{3}{\metre} spacing and \SI{200}{\metre} radius a representative node has
approximately \num{13800} within-row neighbours with deltas $\Delta < 128$
(one byte in LEB128) and approximately \num{133} between-row jumps of roughly
\num{1200} (two bytes), giving approximately \SI{14}{\kilo\byte} per node
compressed versus \SI{56}{\kilo\byte} uncompressed---a \num{4}$\times$
compression ratio.

\paragraph{Encoding.}
The first neighbour index in each row is stored as an absolute unsigned LEB128
varint; subsequent entries encode the non-negative delta from the previous
index.
The \texttt{CompressedCsr} struct stores a \texttt{u64} byte-offset array
(length $N+1$), a \texttt{u32} degree array, and the byte stream.
A lazy \texttt{NeighborIter} decodes varints on demand via a streaming Rust
iterator with zero copying, requiring only two integer additions and two bit
shifts per neighbour.
Because graph traversal is memory-bandwidth-limited, the smaller working set
more than offsets the decode overhead.

\paragraph{Memory management.}
For graphs that fit in RAM the byte stream is heap-allocated; for larger graphs
it is written to an anonymous temporary file during batched parallel
construction (Rayon) and then memory-mapped with \texttt{memmap2}.
In the mmap case the OS page cache manages which pages reside in physical
memory, and peak RSS during construction is bounded by the offset and degree
arrays (approximately \SI{35}{\mega\byte} for \num{2.9} million nodes) plus one
streaming construction batch.
Graphs are persisted in a binary format (\texttt{VGACSR03}) that appends
pre-computed connected-component metadata (Union-Find component IDs and sizes)
so that this information is immediately available on reload without a post-hoc
BFS pass.

\paragraph{Optional Hilbert reordering.}
For unlimited-depth analysis requiring many HyperBall iterations,
an optional Hilbert space-filling curve reordering improves GPU L2 cache
locality.
The reordering maps 2D grid cells to a 1D index where spatially adjacent
cells receive nearby indices, then rebuilds the CSR with remapped and
re-sorted neighbour lists.
Compression is unaffected---the permuted CSR is within
\SI{1}{\percent} of the original size---because Hilbert-ordered neighbours
still produce small deltas.
The inverse permutation (\SI{4}{\byte} per node) is stored in the
\texttt{VGACSR03} file for coordinate restoration.
At depth-limited radii (3--5 iterations), the working set already fits in
GPU L2 cache and Hilbert reordering provides negligible benefit; it is
most useful for large unlimited-depth runs where many iterations amplify
cache miss costs.

\paragraph{GPU streamability.}
A critical property of the level-synchronous HyperBall algorithm is that each
iteration scans every node's compressed neighbour byte-range exactly once in a
predictable sequential pattern.
This allows compressed data to be streamed to the GPU in contiguous batches
with the LEB128 decode performed on-device in CUDA shared memory
(Section~\ref{sec:method:gpu}).
Landmark BFS, by contrast, accesses neighbour lists in frontier-determined
order that depends on graph structure and cannot be pre-scheduled for
sequential streaming; it is therefore confined to the CPU, where the
\texttt{NeighborIter} decoder handles irregular access efficiently.

\begin{figure}[t]
  \centering
  \includegraphics[width=\columnwidth]{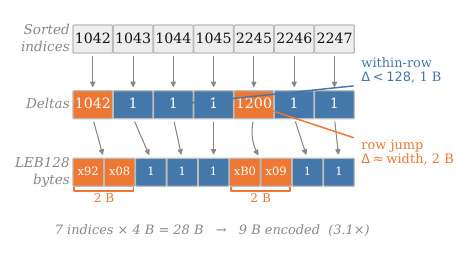}
  \caption{Delta-compressed CSR encoding. Sorted neighbour indices yield
  small deltas that fit in one or two LEB128 bytes. The byte stream is
  memory-mapped for graphs exceeding physical RAM. The sequential structure
  of HyperBall iterations enables the compressed stream to be transferred
  directly to the GPU and decoded on-device.}
  \label{fig:deltacsr}
\end{figure}

\subsection{HyperBall for VGA Metrics}
\label{sec:method:hyperball}

HyperBall~\cite{boldi2011hyperanf,boldi2013hyperball} estimates the
neighbourhood function $|B(v,t)|$---the number of nodes within $t$ hops of
$v$---for all nodes simultaneously by propagating HyperLogLog (HLL) sketches
level-synchronously over the graph.
We adapt it for VGA to compute all BFS-derived metrics in $O(D \cdot |E| \cdot 2^p)$
time, where $D$ is the graph diameter and $p$ the HLL precision, instead of the
$O(N \cdot |E|)$ required by exhaustive BFS.

\paragraph{Initialisation.}
Each node $v$ is assigned an HLL counter comprising $m = 2^p$ registers
(typically $p = 8$ to $14$; we recommend $p = 10$ as the default,
giving $m = 1024$ registers).
Node $v$ is inserted into its own counter using the SplitMix64 finalizer hash,
which matches the hash used in the CUDA kernels (Section~\ref{sec:method:gpu})
for consistent cross-platform behaviour.
Two register arrays are double-buffered in a flat $N \times m/2$ byte layout
(two 4-bit registers per byte) for cache-efficient SIMD access.

\paragraph{Iteration.}
At each step $t$, every node $v$ propagates its neighbourhood estimate to its
neighbours by computing the element-wise maximum (register-wise union) of its
current HLL counter with those of all adjacent nodes:
\begin{equation}
  \mathtt{next}[v][j] = \max\bigl(\mathtt{cur}[v][j],\;
    \max_{w \in \mathcal{N}(v)} \mathtt{cur}[w][j]\bigr),
  \quad j = 0, \ldots, m-1.
  \label{eq:hll-union}
\end{equation}
After the union step the HLL estimator~\cite{flajolet2007hyperloglog} (with
$\alpha_m$ bias correction and small-range linear counting) converts each
register set to a cardinality estimate $\hat{c}_t[v] \approx |B(v,t)|$.

\paragraph{Distance accumulation.}
The increase in estimated neighbourhood size between iterations $t-1$ and $t$
approximates the number of nodes first reached at distance $t$.
Sum-of-distances is accumulated as
\begin{equation}
  \mathtt{sum\_d}[v] \mathrel{+}= t \cdot \bigl(\hat{c}_t[v] - \hat{c}_{t-1}[v]\bigr).
  \label{eq:sumD}
\end{equation}
Visual Mean Depth then follows as
$\mathrm{MD}(v) = \mathtt{sum\_d}[v] / (N_v - 1)$,
where $N_v = |C(v)|$ is the exact component size stored in the VGACSR03 file.
Using exact component sizes for the denominator of integration formulas avoids
amplifying HLL noise through division by an approximate quantity.

\paragraph{Convergence.}
Propagation terminates when no node's cardinality estimate increases by more
than $0.5$ (i.e.\ the rounded change is zero), which occurs after at most $D$
iterations where $D$ is the diameter.
In practice the iteration count equals the graph diameter $D$, which
depends on study area extent, visibility radius, and grid spacing:
bounded-radius configurations at fine spacing converge in fewer than
10 iterations, while city-scale unlimited-radius graphs can require 40
or more.
When a depth limit $d$ is set, HyperBall terminates after exactly $d$
iterations, so the runtime is directly proportional to the depth parameter.
\paragraph{Metric derivation.}
From $\mathrm{MD}(v)$ and exact component size $N_v$, the five BFS-derived
metrics are computed in closed form: Integration [HH] as the reciprocal of
Real Relative Asymmetry (RRA)~\cite{hillier1984social};
Integration [Tekl] as $\log_2(({\rm MD}+2)/3)$;
Integration [P-value] as $\max(0, 1 - \mathrm{RA})$~\cite{turner2004depthmap}
where $\mathrm{RA} = 2(\mathrm{MD}-1)/(N_v-2)$; and Point First Moment as
$\mathrm{MD}(v) \times \deg(v)$.
Local metrics (Connectivity, Control, Controllability, Clustering,
Point Second Moment) are computed exactly from the 1-hop neighbourhood and are
unaffected by the HLL approximation.
Entropy and Relativised Entropy require the full depth distribution and are
returned as NaN, consistent with landmark BFS.

\begin{algorithm}[t]
  \caption{HyperBall for VGA metrics}
  \label{alg:hyperball}
  \begin{algorithmic}[1]
    \REQUIRE Graph $G = (V, E)$ in delta-compressed CSR; precision $p$;
             exact component sizes $\{N_v\}$
    \STATE $m \leftarrow 2^p$; allocate $\mathtt{cur}[N \times m/2]$,
           $\mathtt{next}[N \times m/2]$ (zero-initialised; 4-bit packed)
    \FOR{each $v \in V$}
      \STATE Insert $v$ into $\mathtt{cur}[v]$ using SplitMix64 hash
    \ENDFOR
    \STATE Estimate initial cardinalities $\hat{c}_0[v]$ for all $v$
    \FOR{$t = 1, 2, \ldots$}
      \FOR{each $v \in V$ \textbf{in parallel}}
        \STATE $\mathtt{next}[v] \leftarrow \mathtt{cur}[v]$
        \FOR{each $w \in \mathcal{N}(v)$}
          \STATE $\mathtt{next}[v] \leftarrow \max(\mathtt{next}[v],
                 \mathtt{cur}[w])$ \COMMENT{register-wise}
        \ENDFOR
      \ENDFOR
      \STATE Estimate $\hat{c}_t[v]$ for all $v$ \COMMENT{HLL estimator}
      \STATE $\mathtt{sum\_d}[v] \mathrel{+}=
             t \cdot (\hat{c}_t[v] - \hat{c}_{t-1}[v])$ for all $v$
      \IF{$\max_v (\hat{c}_t[v] - \hat{c}_{t-1}[v]) \leq 0.5$}
        \STATE \textbf{break} \COMMENT{converged}
      \ENDIF
      \STATE swap $\mathtt{cur} \leftrightarrow \mathtt{next}$
    \ENDFOR
    \STATE Derive VGA metrics from $\mathtt{sum\_d}[v]$ and exact $N_v$
  \end{algorithmic}
\end{algorithm}

\subsection{GPU Acceleration}
\label{sec:method:gpu}

The CPU HyperBall implementation (Section~\ref{sec:method:hyperball}) is
bottlenecked by DRAM bandwidth: each iteration reads every node's $m$-register
counter and those of all its neighbours from main memory.
At \num{2.9} million nodes with $p = 8$ ($m = 256$ registers, packed to
\SI{128}{\byte} per node) and \num{60} billion edges, each iteration performs
approximately \SI{7.7}{\tera\byte} of random register reads at
\SI{40}{\giga\byte\per\second} DDR4 bandwidth, yielding approximately
\SI{190}{\second} per iteration and \SI{1.6}{\hour} total for $D \approx 30$
iterations.

The GPU solution moves the HLL registers to device memory, where the RTX
3080 Ti Laptop GPU's \SI{512}{\giga\byte\per\second} peak GDDR6 bandwidth
largely eliminates the register-access bottleneck.
When the compressed graph exceeds \SI{16}{\giga\byte} VRAM it is streamed from
host memory to the device in batches per iteration, with LEB128 decoding
performed on-device.
This is the critical enabling property of the delta-compressed CSR format: the
sequential scan pattern of HyperBall allows contiguous byte ranges to be
transferred and decoded without frontier-dependent random access.

\begin{figure}[t]
  \centering
  \includegraphics[width=\columnwidth]{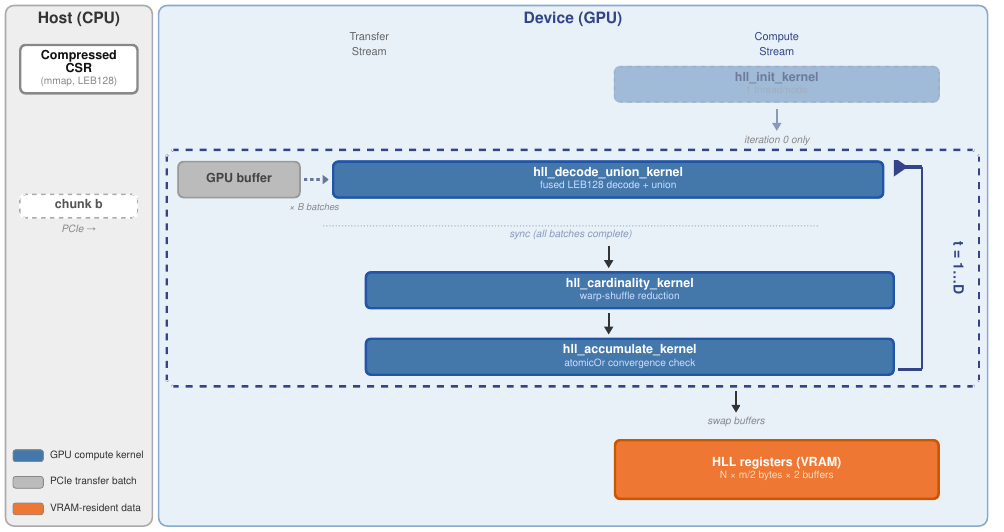}
  \caption{GPU HyperBall execution pipeline. HLL register arrays reside in
  VRAM throughout. Compressed graph data is streamed to the device in batches
  of \num{10000} nodes via a transfer stream, overlapping with kernel execution
  on the compute stream. The fused decode-union kernel decodes LEB128 varints
  in shared memory and performs register max-union in a single pass.}
  \label{fig:gpupipeline}
\end{figure}

\paragraph{CUDA kernels.}
Four kernels implement the GPU HyperBall pipeline
(Figure~\ref{fig:gpupipeline}):
\begin{enumerate}
\item \textbf{hll\_init\_kernel} (one thread per node): zeros all registers
and inserts each node into its own counter using the SplitMix64 hash,
identical to the Rust CPU implementation.

\item \textbf{hll\_decode\_union\_kernel} (one block per node):
thread~0 decodes the LEB128 delta stream for its node's neighbours into a
\num{4096}-entry shared-memory buffer.
All threads then stride over the $m$ HLL registers, computing the element-wise
maximum against the decoded neighbours' registers in device memory.
Nodes with more than \num{4096} neighbours are handled in chunks with
synchronisation between each chunk.

\item \textbf{hll\_cardinality\_kernel} (one block per node): threads stride
over the $m$ registers accumulating the harmonic-mean sum; warp-level
shuffle reduction followed by shared-memory cross-warp reduction yields
the per-node cardinality estimate, applying HLL++ bias correction
and small-range linear counting.

\item \textbf{hll\_accumulate\_kernel} (one thread per node): accumulates
the distance sum via Equation~\eqref{eq:sumD} and sets an atomic convergence
flag (\texttt{atomicOr}) if any node's cardinality increased by more than
$0.5$.
\end{enumerate}

All kernels use 4-bit packed register format (two registers per byte) and
support precisions up to $p = 16$ via register striding with block sizes up to
\num{1024}.
When optional Hilbert reordering is enabled (VGACSR03 format), the
decode-union kernel benefits from improved L2 cache locality as consecutive
blocks access overlapping neighbour sets; this is most beneficial for
unlimited-depth runs with many iterations.

\paragraph{Dual-stream execution.}
A second CUDA stream handles host-to-device transfers of compressed graph
batches, overlapping PCIe data movement with kernel execution on the compute
stream.
The HLL register arrays ($N \times m/2$ bytes $\times$ 2 buffers) remain
resident in VRAM throughout all iterations.

\section{Experimental Evaluation}
\label{sec:evaluation}

\subsection{Experimental Setup}
\label{sec:evaluation:setup}

All experiments run on a laptop workstation with an Intel i7-12700H (20 logical
cores), \SI{32}{\giga\byte} DDR4 RAM, and an NVIDIA RTX 3080 Ti Laptop GPU
(\SI{16}{\giga\byte} GDDR6, 256-bit bus, PCIe 4.0 x8), running CUDA 12.0
(sm\_86).

The study area is Valdivia, Chile (EPSG:32718), with building footprints from
OpenStreetMap and a municipal boundary polygon.
To evaluate both accuracy and scaling, we construct circular sub-regions of
increasing radius (\SIlist{200;300;500;750;1000}{\metre}) centred on the city
centre, with grid spacings of \SIlist{3;5;7;10;20}{\metre} and unlimited
visibility radius, producing problem sizes ranging from \num{235} to
\num{236000} grid cells.
All runs use the sparkSieve2 visibility backend
(Section~\ref{sec:method:visibility}), which produces identical edge sets to
depthmapX, isolating accuracy differences to the HLL approximation alone.
Both tools are run at topological depth limit~3 (the standard local
measure in VGA practice~\cite{turner2004depthmap,koutsolampros2019dissecting}).

The baseline is depthmapXcli~\cite{varoudis2012depthmapx} (version 0.8.0),
run with a \SI{30}{\minute} timeout per configuration.
Our tool is tested at HLL precisions $p \in \{8, 10, 12\}$, with
$p = 10$ as the recommended default balancing accuracy and speed.

\subsection{Accuracy Validation}
\label{sec:evaluation:accuracy}

We validate HyperBall metric estimates against depthmapX across 20 matched
configurations (all area/spacing combinations where depthmapX completed within
the timeout and both tools produced spatially coincident grid points, all at
depth limit~3).
For each configuration, we spatially join the output point sets and compute
Pearson~$r$ on Visual Mean Depth and Spearman~$\rho$ on Integration~[HH].
We use Spearman rank correlation for Integration~[HH] because the
Hillier--Hanson normalisation applies a nonlinear transform that amplifies
small mean-depth errors into large absolute integration differences,
particularly for nodes near the distribution extremes; $R^2$ in the original
scale is unreliable as an accuracy metric for this quantity.

Table~\ref{tab:accuracy} summarises accuracy across HLL precisions.
At $p = 10$ (the recommended default), Mean Depth achieves Pearson
$r = 0.999$ with median relative error \SI{1.7}{\percent}, and
Integration~[HH] achieves Spearman $\rho = 0.893$ on average.
Increasing to $p = 12$ reduces median error to \SI{0.8}{\percent} with
$\rho = 0.964$, at the cost of ${\sim}4{\times}$ longer BFS time
(Section~\ref{sec:evaluation:performance}).
At depth limit~3, Mean Depth values cluster in a narrow range (most
nodes are reached within 3 hops in high-connectivity visibility graphs), so
Pearson~$r$ is more sensitive to small absolute errors than in
unlimited-depth configurations; median relative error is a more stable
accuracy indicator.

\begin{table}[t]
  \centering
  \caption{HyperBall accuracy vs depthmapX across HLL precisions
  (depth limit~3).
  Mean Depth uses Pearson $r$; Integration [HH] uses Spearman $\rho$.
  Values are averages over 20 matched configurations (\num{235}--\num{42705}
  cells). Ranges in parentheses.}
  \label{tab:accuracy}
  \small
  \begin{tabular}{@{}ccccc@{}}
    \toprule
    $p$ & MD $r$ (range) & MD med.\ err & IHH $\rho$ (range) & $n$ \\
    \midrule
    8  & 0.996 (0.983--0.999) & 4.0\% & 0.789 (0.301--0.974) & 20 \\
    10 & 0.999 (0.998--1.000) & 1.7\% & 0.893 (0.708--0.994) & 20 \\
    12 & 1.000 (0.999--1.000) & 0.8\% & 0.964 (0.882--0.994) & 20 \\
    \bottomrule
  \end{tabular}
\end{table}

\begin{figure}[t]
  \centering
  \begin{subfigure}[b]{0.48\columnwidth}
    \includegraphics[width=\textwidth]{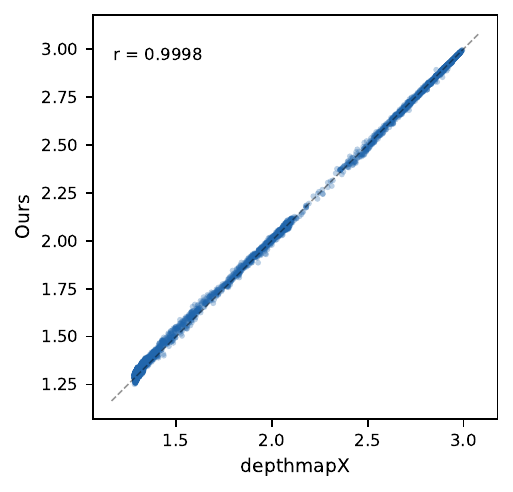}
    \caption{Visual Mean Depth}
  \end{subfigure}
  \hfill
  \begin{subfigure}[b]{0.48\columnwidth}
    \includegraphics[width=\textwidth]{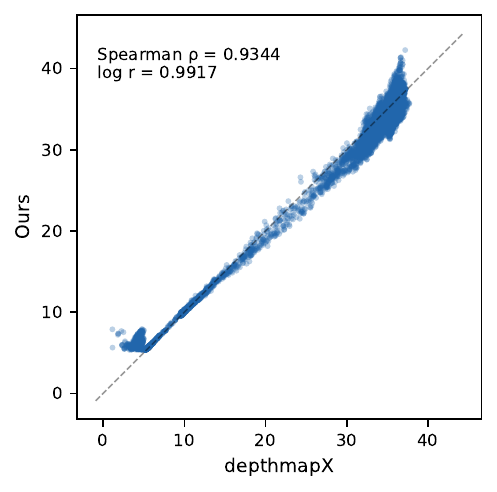}
    \caption{Integration [HH]}
  \end{subfigure}
  \caption{Our tool ($p = 10$, depth limit~3) vs depthmapX for a
  representative configuration.
  Mean Depth (left) annotated with Pearson $r$; Integration [HH] (right)
  annotated with Spearman $\rho$.}
  \label{fig:accuracy}
\end{figure}

\begin{figure}[t]
  \centering
  \includegraphics[width=\columnwidth]{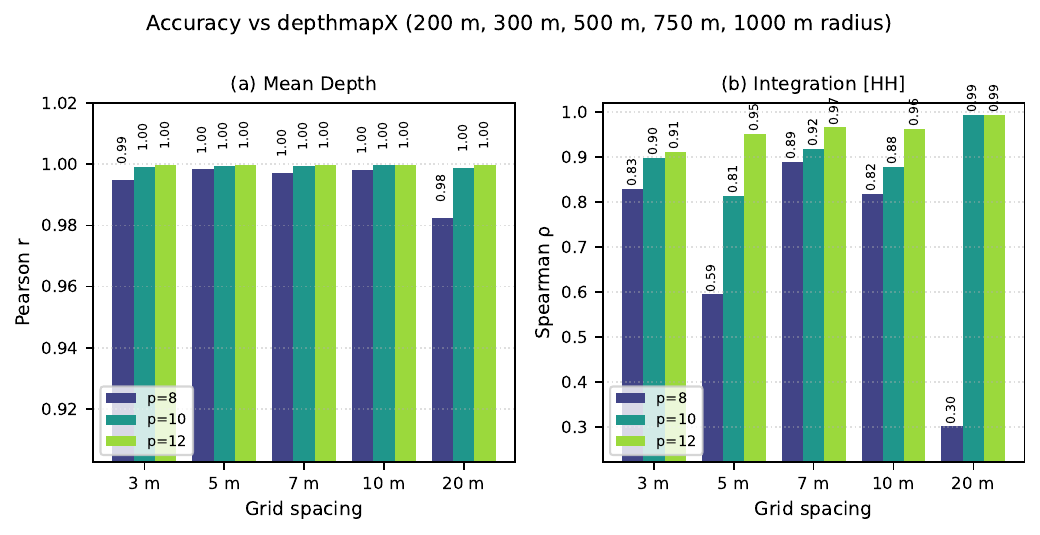}
  \caption{Accuracy heatmaps by area radius and grid spacing.
  Top row: Pearson $r$ for Mean Depth; bottom row: Spearman $\rho$ for
  Integration [HH].
  Accuracy improves monotonically with precision and problem size.}
  \label{fig:accuracy-heatmap}
\end{figure}

Figure~\ref{fig:accuracy} shows representative per-node scatter plots for
the validation configuration.
Local metrics (connectivity, control, controllability, clustering coefficient,
point second moment) are computed exactly from the 1-hop neighbourhood and are
unaffected by the HLL approximation.
Accuracy improves monotonically with precision; it also varies with problem
geometry, as larger study areas contain more spatially diverse subgraphs that
provide better averaging of HLL noise
(Figure~\ref{fig:accuracy-heatmap}).

\subsection{Performance and Scaling}
\label{sec:evaluation:performance}

\begin{figure}[t]
  \centering
  \includegraphics[width=\columnwidth]{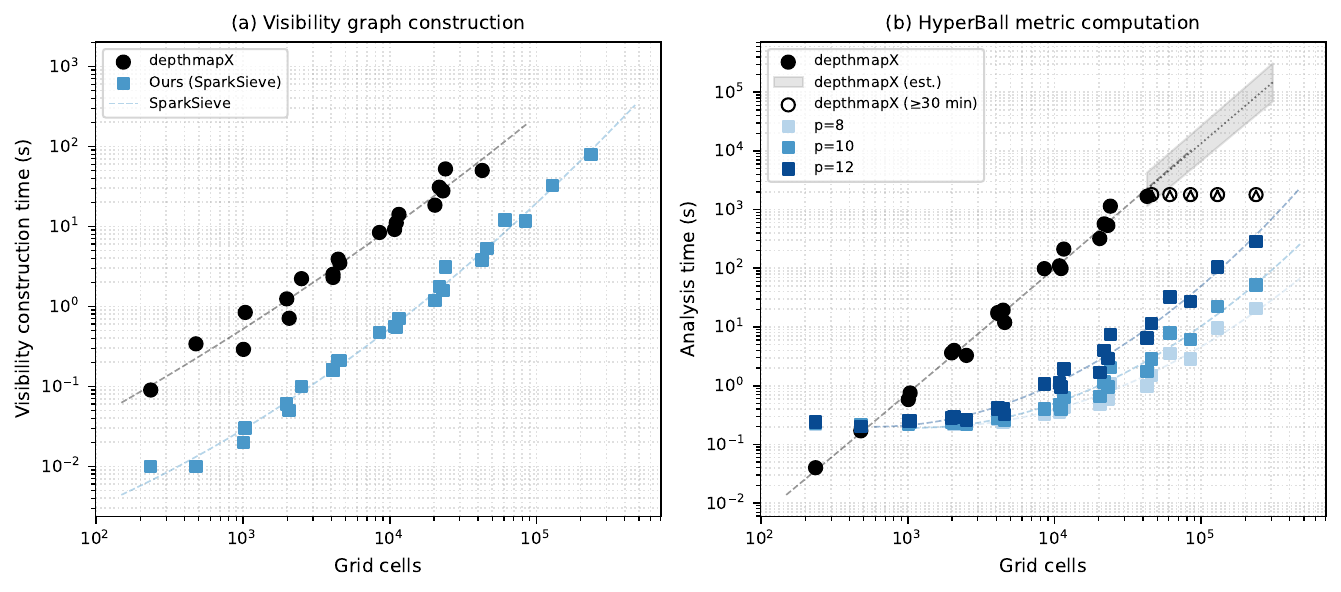}
  \caption{Log-log scaling by grid cells.
  (a)~Visibility graph construction: our SparkSieve implementation compared
  with depthmapX.  Both use the same algorithm; our parallelised Rust port
  is consistently faster.
  (b)~HyperBall metric computation by HLL precision: $p = 8$ and $p = 10$
  remain well below depthmapX at all sizes; $p = 12$ approaches depthmapX
  cost at scale.  Dashed curves are quadratic fits in log-log space.}
  \label{fig:scaling}
\end{figure}

\begin{figure}[t]
  \centering
  \includegraphics[width=\columnwidth]{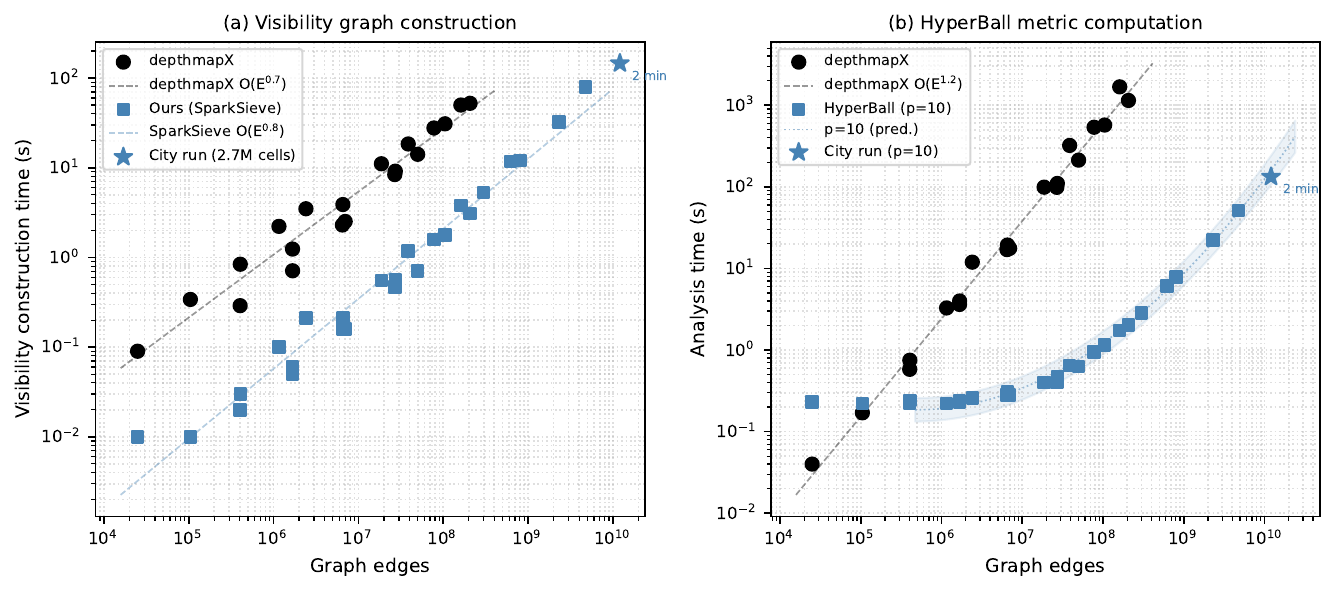}
  \caption{Log-log scaling by graph edges.  Plotting time against $|E|$
  collapses the radius/spacing scatter visible in Figure~\ref{fig:scaling},
  because runtime depends on edge count, not cell count alone.
  Shaded bands are 95\% prediction intervals from quadratic log-log fits.
  Star markers show the city-scale run
  (\S\ref{sec:casestudy}; \num{2.7e6}~cells, \num{12.1e9}~edges),
  which falls within both prediction bands.}
  \label{fig:scaling-edges}
\end{figure}

Figure~\ref{fig:scaling} shows wall-clock time versus grid cells on a
log-log scale, with visibility construction (Figure~\ref{fig:scaling}a)
and HyperBall metric computation (Figure~\ref{fig:scaling}b) plotted
separately for each HLL precision.
Because configurations with the same cell count but different
radius/spacing combinations produce different edge counts, the
cell-based view exhibits scatter at each cell count.
Figure~\ref{fig:scaling-edges} re-plots the same data against edge count
$|E|$, which collapses this scatter into coherent scaling trends and
enables direct comparison with the city-scale run
(\S\ref{sec:casestudy}).
depthmapX completed 20 of the 25 configurations within the
\SI{30}{\minute} timeout; its largest completed run was \num{42705} cells
(\SI{1000}{\metre} radius, \SI{7}{\metre} spacing) in \SI{1732}{\second}
(${\sim}$\SI{29}{\minute}).

\begin{figure}[t]
  \centering
  \includegraphics[width=\columnwidth]{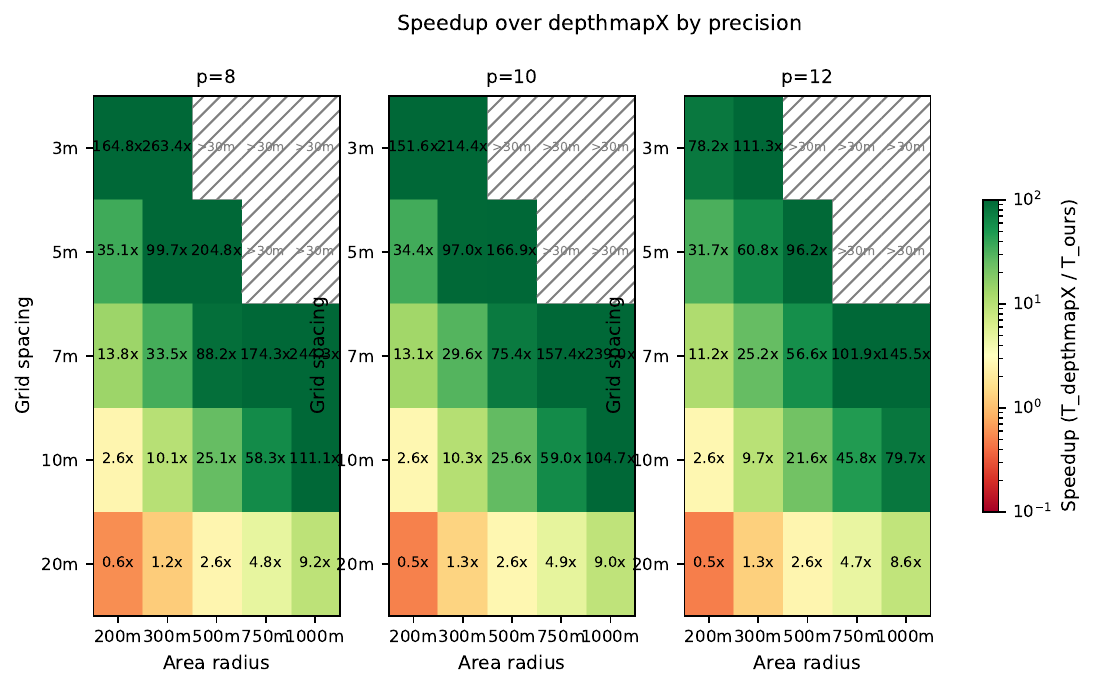}
  \caption{Speedup of our tool over depthmapX ($T_{\text{dm}} / T_{\text{ours}}$)
  by area radius and grid spacing, at four HLL precisions.
  Hatched cells indicate depthmapX timeouts (no reference time available).
  Values below $1{\times}$ at coarse spacings reflect GPU initialisation
  overhead dominating small graphs (see crossover discussion in text).}
  \label{fig:speedup}
\end{figure}

Speedup grows rapidly with problem size: $34{\times}$ at \num{4100} cells,
$152{\times}$ at \num{11500} cells, and $214{\times}$ at \num{24000} cells
(all at $p = 10$; Figure~\ref{fig:speedup}).
At the largest matched configuration (\num{42705} cells, \num{161}~million
edges), our tool completes in \SI{7.25}{\second} versus depthmapX's
\SI{1732}{\second}---a $239{\times}$ speedup.
Beyond depthmapX's timeout, our tool handles \num{236000} cells
(\SI{1000}{\metre} radius, \SI{3}{\metre} spacing, \num{4.8} billion edges)
in \SI{137}{\second} at $p = 10$.

\paragraph{Edge density and superlinear scaling.}
Both tools exhibit superlinear scaling because unlimited visibility radius
causes edge density to grow with problem size.
Table~\ref{tab:edge-density} quantifies this: average degree rises from
\num{106} at \num{235} cells to \num{20237} at \num{236000} cells, meaning
$|E|$ grows nearly quadratically (${\sim}N^{1.9}$) with $N$ under
unlimited visibility.
HyperBall's theoretical cost is $O(D \times |E| \times 2^p)$: with
$|E| \propto N^2$ the analysis phase becomes the bottleneck at high precision.
In practice, most urban VGA studies use a bounded visibility radius
(\SIrange{100}{400}{\metre}), which caps average degree and
restores $O(N)$ edge growth---the superlinear tail is an artefact of the
unbounded benchmark configuration, not the algorithm.

\begin{table}[t]
\centering
\caption{Edge density growth under unlimited visibility.
Average degree rises with problem size because larger study areas expose
more inter-visible cell pairs.  With a bounded visibility radius,
average degree stabilises and $|E|$ grows linearly with $N$.}
\label{tab:edge-density}
\small
\begin{tabular}{@{}r r r l@{}}
\toprule
{Cells} & {Edges} & {Avg.\ degree} & {Config} \\
\midrule
\num{235}    & \num{25}K    & \num{106}    & 200\,m / 20\,m \\
\num{1007}   & \num{404}K   & \num{401}    & 200\,m / 10\,m \\
\num{4106}   & \num{6.5}M   & \num{1576}   & 200\,m / 5\,m  \\
\num{11555}  & \num{50}M    & \num{4321}   & 200\,m / 3\,m  \\
\num{42705}  & \num{161}M   & \num{3775}   & 1000\,m / 7\,m \\
\num{61005}  & \num{808}M   & \num{13237}  & 500\,m / 3\,m  \\
\num{128943} & \num{2.3}B   & \num{17836}  & 750\,m / 3\,m  \\
\num{235983} & \num{4.8}B   & \num{20237}  & 1000\,m / 3\,m \\
\bottomrule
\end{tabular}
\end{table}

\paragraph{Pipeline phase breakdown.}
Total runtime comprises three phases: grid generation (constant per area),
visibility graph construction, and GPU HyperBall BFS.
At depth limit~3, the BFS and visibility phases are roughly balanced:
BFS accounts for \SIrange{39}{47}{\percent} of combined phase time across
configurations at $p = 10$ (Table~\ref{tab:breakdown}).
At $p = 8$, BFS drops to \SIrange{21}{35}{\percent};
at $p = 12$, BFS dominates at large scale (\SI{78}{\percent} at
\num{236000} cells) as the $4{\times}$ register count increase outweighs
the visibility phase.

\begin{table}[t]
\centering
\caption{Pipeline phase breakdown: visibility graph construction
(\textsc{vis}) versus GPU HyperBall BFS (\textsc{bfs}) time in seconds,
with BFS share of total time in parentheses.
Grid generation time is omitted (${<}\SI{2}{\second}$ in all cases).
All runs at depth limit~3.}
\label{tab:breakdown}
\small
\begin{tabular}{@{}r r l r r r r r r@{}}
\toprule
& & & \multicolumn{2}{c}{$p = 8$} & \multicolumn{2}{c}{$p = 10$} & \multicolumn{2}{c}{$p = 12$} \\
\cmidrule(lr){4-5} \cmidrule(lr){6-7} \cmidrule(lr){8-9}
{Cells} & {Edges} & {Config} & {\textsc{vis}} & {\textsc{bfs}} & {\textsc{vis}} & {\textsc{bfs}} & {\textsc{vis}} & {\textsc{bfs}} \\
\midrule
\num{11555}  & \num{50}M   & 200\,m / 3\,m  & 0.8 & 0.4\,(35\%) & 0.7 & 0.6\,(47\%) & 0.8 & 1.9\,(72\%) \\
\num{23991}  & \num{207}M  & 300\,m / 3\,m  & 3.0 & 1.1\,(26\%) & 3.1 & 2.0\,(40\%) & 3.0 & 7.4\,(71\%) \\
\num{61005}  & \num{808}M  & 500\,m / 3\,m  & 11.9 & 3.5\,(23\%) & 12.1 & 7.9\,(39\%) & 11.8 & 32.2\,(73\%) \\
\num{128943} & \num{2.3}B  & 750\,m / 3\,m  & 32.2 & 9.6\,(23\%) & 32.3 & 22.2\,(41\%) & 31.4 & 104\,(77\%) \\
\num{235983} & \num{4.8}B  & 1000\,m / 3\,m & 78.6 & 20.6\,(21\%) & 79.5 & 51.6\,(39\%) & 81.3 & 286\,(78\%) \\
\bottomrule
\end{tabular}
\end{table}

\begin{figure}[t]
  \centering
  \includegraphics[width=\columnwidth]{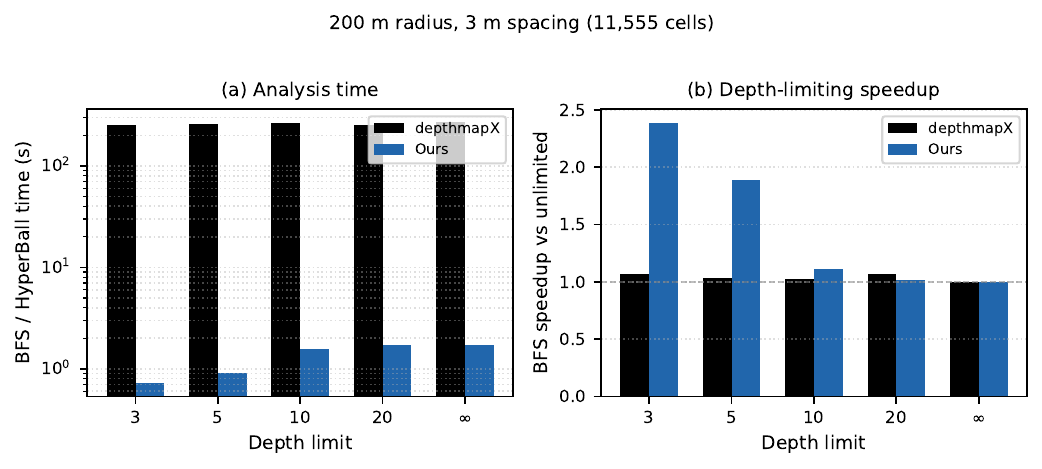}
  \caption{Depth-limited BFS comparison at \num{11555} cells ($p = 10$).
  (a)~Absolute BFS time: depthmapX shows no reduction with
  decreasing depth, while HyperBall cost scales linearly with iteration count.
  (b)~Speedup relative to each tool's own unlimited-depth time:
  our depth-3 BFS is $2.4{\times}$ faster than our unlimited run, while
  depthmapX shows negligible benefit from depth limiting.}
  \label{fig:depth-bfs}
\end{figure}

\paragraph{Depth-limited analysis.}
Both depthmapX and our tool support topological depth limits
(called radius-$n$ in axial analysis~\cite{hillier1984social}),
restricting BFS to a maximum of $n$ hops from each source node.
Depth-3 local integration is the canonical local VGA
measure~\cite{turner2004depthmap,koutsolampros2019dissecting}, widely
used for predicting pedestrian movement at neighbourhood
scale~\cite{hillier1993natural,hillier1996space}.
Global (unlimited-depth) integration captures macro-scale structure but is
sensitive to study area boundary definition~\cite{gil2017edgeeffects}---a
limitation noted by Ericson et al.~\cite{ericson2021robustness}, who showed
that VGA metrics vary with both resolution and boundary placement.
Table~\ref{tab:depth-sweep} compares BFS time for both tools across depth
limits at \num{11555} cells (\SI{200}{\metre} radius, \SI{3}{\metre} spacing,
\num{50}M edges).
depthmapX's BFS time shows no substantial reduction with depth limiting at
this configuration (Figure~\ref{fig:depth-bfs}a).
Inspection of the depthmapX source code
(\texttt{vgavisualglobal.cpp}) confirms that the BFS frontier \emph{is}
correctly pruned: nodes beyond the depth limit are counted but not expanded.
The flat timing arises instead from the topology of visibility graphs.
With unlimited visibility radius and \SI{3}{\metre} spacing, each node
sees hundreds of others at hop distance~1 (average degree \num{4321} at
this configuration; Table~\ref{tab:edge-density}).
The graph diameter is consequently very small---typically 3--6 hops---so a
depth-3 BFS already reaches the vast majority of reachable nodes and
performs nearly the same work as an unlimited traversal.
Additionally, depthmapX's per-source BFS clears a dense visited-flag matrix
covering the full grid ($\text{rows} \times \text{cols}$) before each
source node, imposing a fixed $O(G)$ overhead per source regardless of
how many nodes the BFS actually visits.

\begin{figure}[t]
  \centering
  \includegraphics[width=\columnwidth]{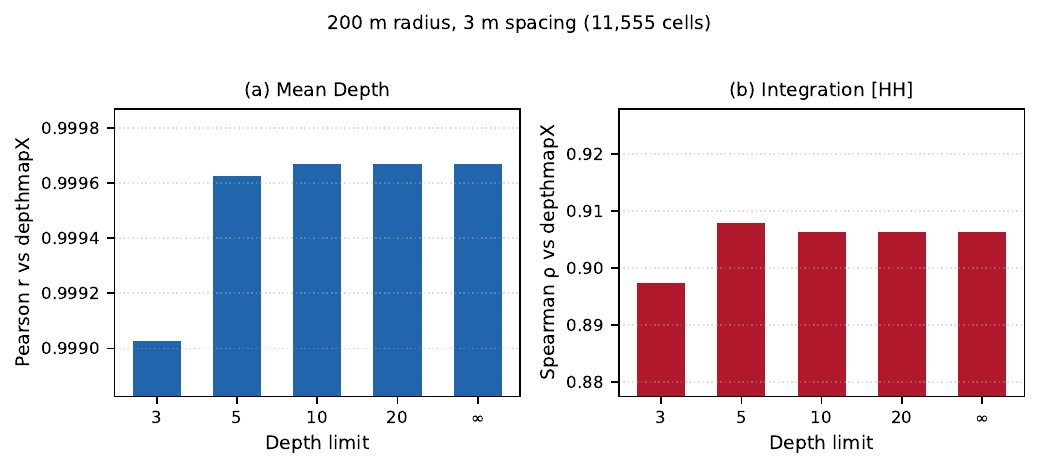}
  \caption{Accuracy across depth limits at \num{11555} cells ($p = 10$,
  depths 3--unlimited).
  (a)~Mean Depth Pearson $r \geq 0.999$ at all depths.
  (b)~Integration~[HH] Pearson $r$ remains above 0.97 at all tested
  depths.}
  \label{fig:depth-accuracy}
\end{figure}

HyperBall, by contrast, converges in exactly $\min(d, D)$ iterations where
$d$ is the depth limit and $D$ the diameter.
At depth limit~3, our BFS completes in \SI{0.72}{\second} versus depthmapX's
\SI{253}{\second}---a $352{\times}$ BFS-phase speedup
(Figure~\ref{fig:depth-bfs}; end-to-end speedup is lower as visibility
construction cost is shared across depth settings).
With early stopping enabled, unlimited-depth HyperBall converges at the
graph diameter in \SI{1.72}{\second}---$2.4{\times}$ slower than depth-3
but still $157{\times}$ faster than depthmapX.
The architectural difference is fundamental: per-source BFS cost depends on
the number of \emph{nodes visited}, which plateaus rapidly in
high-connectivity graphs; HyperBall cost depends on the number of
\emph{iterations}, which scales linearly with the depth limit regardless of
graph connectivity.
BFS time increases monotonically with depth setting (Table~\ref{tab:depth-sweep}):
\SI{0.72}{\second} at $d = 3$, \SI{0.91}{\second} at $d = 5$,
\SI{1.70}{\second} at $d = 20$---while depthmapX remains at
\SIrange{252}{270}{\second} throughout.
Since practitioners routinely use radius-3 or radius-5 for local VGA, this
depth-proportional speedup applies to the most common analytical workflow.
Mean Depth accuracy at $p = 10$ is Pearson $r \geq 0.999$ across all
tested depth settings including unlimited
(Figure~\ref{fig:depth-accuracy}), confirming that the SparkSieve
visibility port produces identical edge sets and the HLL approximation
introduces minimal error regardless of convergence depth.

\begin{table}[t]
\centering
\caption{Depth-limited BFS comparison at \num{11555} cells (\SI{200}{\metre}
radius, \SI{3}{\metre} spacing, \num{50}M edges, $p = 10$).
depthmapX BFS time varies little across depth settings;
HyperBall converges in $\min(d, D)$ iterations, directly exploiting
the reduced depth.}
\label{tab:depth-sweep}
\small
\begin{tabular}{@{}l r r r c@{}}
\toprule
{Depth} & {dmX BFS (s)} & {Ours BFS (s)} & {BFS Speedup} & {MD $r$} \\
\midrule
Unlimited & 270.3 & 1.72 & $157{\times}$ & 1.000 \\
20        & 252.3 & 1.70 & $148{\times}$ & 1.000 \\
10        & 263.5 & 1.55 & $170{\times}$ & 1.000 \\
5         & 261.1 & 0.91 & $287{\times}$ & 1.000 \\
3         & 253.4 & 0.72 & $352{\times}$ & 0.999 \\
\bottomrule
\end{tabular}
\end{table}

\section{Case Study: Valdivia, Chile}
\label{sec:casestudy}

To demonstrate city-scale applicability, we ran our full pipeline on
the \SI{49.45}{\kilo\metre\squared} Valdivia study area at
\SI{5}{\metre} grid spacing with a \SI{400}{\metre} visibility radius
and depth limit~3, producing a graph of \num{2706968} nodes and
\num{12.1e9} edges.
The complete analysis---grid generation, SparkSieve visibility
construction, delta-compressed CSR serialisation, and GPU HyperBall
metric computation at $p = 10$---completed in \SI{333}{\second}
(${\sim}$\SI{5.5}{\minute}) on a single laptop GPU
(RTX~3080~Ti, \SI{16}{\giga\byte}).

\begin{figure}[t]
  \centering
  \includegraphics[width=\columnwidth]{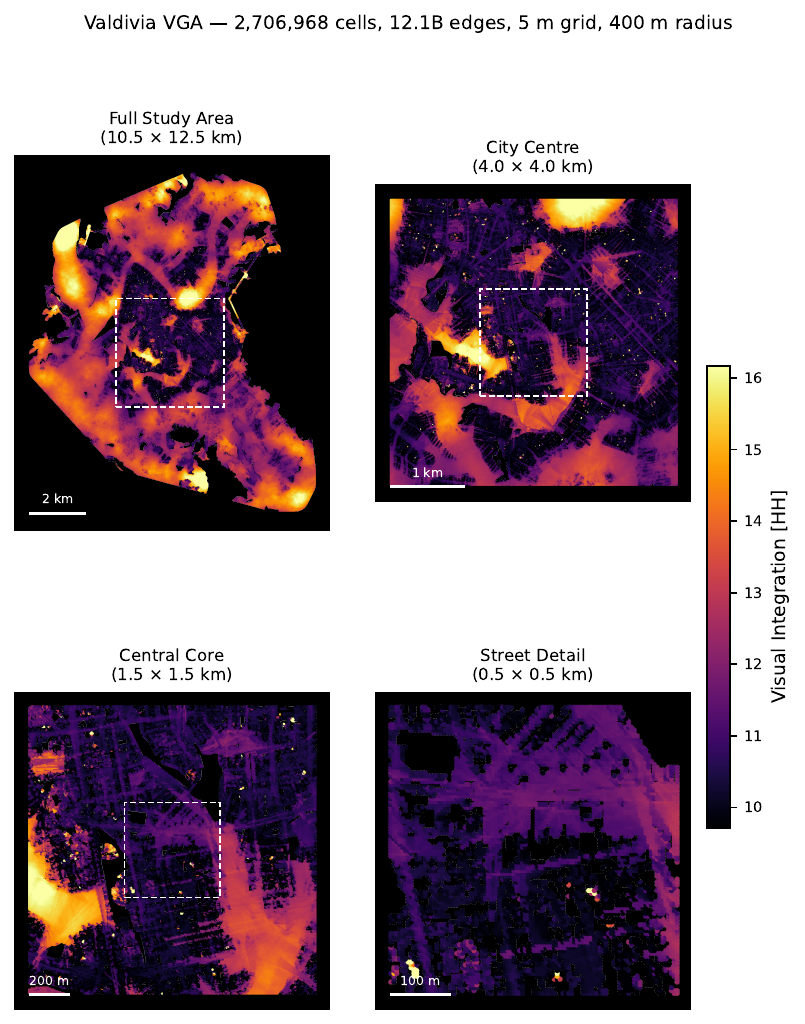}
  \caption{Visual Integration [HH] for Valdivia at \SI{5}{\metre}
  spacing with \SI{400}{\metre} radius (\num{2.7e6} cells,
  \num{12.1e9} edges).  Successive panels zoom from the full
  \SI{49.45}{\kilo\metre\squared} study area to a
  \SI{500}{\metre}~$\times$~\SI{500}{\metre} street-level detail.
  Dashed boxes indicate the extent of the next zoom level.}
  \label{fig:valdivia}
\end{figure}

Figure~\ref{fig:valdivia} shows Visual Integration [HH] at four zoom
levels.  At the city scale, the commercial core along Avenida Picarte
and the central grid around Plaza de la Rep\'{u}blica exhibit the
highest integration, consistent with prior neighbourhood vitality
analysis by Zumelzu \& Barrientos-Trinanes~\cite{zumelzu2019valdivia}.
The river corridor and peripheral residential areas show lower values.
At street level, the \SI{5}{\metre} resolution reveals fine-grained
variation: individual building footprints create local shadows in the
integration field, and through-block passages appear as narrow
high-integration corridors.

This configuration is far beyond the practical reach of exact BFS tools
such as depthmapX, which timed out on graphs exceeding \num{42705} cells
in our benchmarks (\S\ref{sec:evaluation:performance}).
The two dominant phases---visibility construction (\SI{146}{\second}) and
GPU HyperBall BFS (\SI{133}{\second})---account for \SI{279}{\second} of
the \SI{333}{\second} total, with the remainder spent on grid generation
(\SI{32}{\second}), obstacle rasterisation (\SI{11}{\second}), and
GeoPackage output (\SI{2}{\second}).
Both dominant phases fall within the 95\% prediction bands extrapolated
from the benchmark surface (Figure~\ref{fig:scaling-edges}), confirming
that performance scales predictably to city-scale graphs.

\section{Discussion}
\label{sec:discussion}

\paragraph{Accuracy of HLL-based metrics.}
The HyperLogLog approximation introduces a precision-dependent bias in
BFS-derived metrics.
At $p = 10$, the standard error on raw cardinality is
$1.04/\sqrt{2^{10}} \approx \SI{3.2}{\percent}$, but per-iteration HLL
noise partially cancels in the distance sum, yielding empirical median
relative errors of \SI{1.7}{\percent} on Mean Depth
(Table~\ref{tab:accuracy}).
The Hillier--Hanson Integration normalisation amplifies errors for nodes near
the distribution extremes, making $R^2$ in the original scale an unreliable
accuracy metric.
We recommend reporting Spearman $\rho$ or log-space Pearson $r$ for
Integration~[HH], and note that the Teklenburg formulation is more
robust to approximation error due to its logarithmic transform.

\paragraph{Resolution sensitivity.}
Ericson et al.~\cite{ericson2021robustness} showed that VGA metrics vary with
grid spacing.
Our system's speed makes multi-resolution analysis practical: a full sweep
across five spacings and five area sizes at $p = 10$ completes in under
\SI{5}{\minute} (Table~\ref{tab:breakdown}), enabling practitioners to assess
robustness without the days-long depthmapX runs previously required.

\paragraph{Depth-proportional speedup as a practical advantage.}
The most significant practical consequence of HyperBall for VGA is that
computation time scales with topological depth.
In standard VGA practice, local integration at depth-3 or depth-5
is the primary analytical tool for neighbourhood-scale studies, while global
(unlimited-depth) metrics are used selectively and are known to be sensitive
to study area boundaries~\cite{hillier1996space,gil2017edgeeffects}.
depthmapX's per-source BFS correctly prunes the frontier at the depth
limit, but this yields negligible speedup because visibility graphs have
very small diameters (typically 3--6 hops): high node connectivity means
that even a depth-3 BFS visits nearly every reachable node, performing
essentially the same work as an unlimited traversal.
This is not a missed optimisation in depthmapX---it is an inherent property
of per-source BFS on high-connectivity graphs.
Moreover, depthmapX's per-source BFS is the natural architecture for a
general-purpose space syntax platform that also supports axial, segment,
and angular analysis, all of which require exact per-node distances or
full depth distributions that HyperBall cannot provide (our system returns
NaN for entropy and relativised entropy).
The tradeoff we make---approximate cardinalities in place of exact
distances---is specifically suited to VGA's distance-sum metrics.
Our system sidesteps the frontier plateau entirely: HyperBall's cost depends on
iteration count, not nodes visited, so converging in $\min(d, D)$ iterations
gives a speedup proportional to $D/d$ regardless of graph connectivity.
At depth-3 this yields a $352{\times}$ BFS-phase speedup at the tested
configuration---the depth-3 analysis that practitioners already prefer is
also the fastest to compute.
At unlimited depth, HyperBall converges at the graph diameter
(typically 3--6 iterations) in \SI{1.72}{\second}---still
$157{\times}$ faster than depthmapX, while depth-3 provides an additional
$2.4{\times}$ speedup over unlimited.
This alignment between analytical best practice and computational efficiency
means that city-scale local VGA becomes routine rather than exceptional.
For the less common case of unlimited-depth global analysis, a bounded
visibility radius (typically \SIrange{100}{400}{\metre}) caps edge density
and graph diameter, keeping total work
manageable~\cite{turner2004depthmap,koutsolampros2019dissecting}.

\paragraph{Limitations.}
The current implementation uses 2D line-of-sight visibility with no terrain
model; hilly cities would require a 2.5D or 3D extension.
Vegetation is included as a sightline obstruction but not as a spatial barrier
(pedestrians can walk through vegetated areas).
The GPU pipeline requires CUDA-capable hardware; without a GPU, the CPU
HyperBall fallback is available but substantially slower.

\paragraph{GPU scaling ceiling.}
The quadratic log-log fits in Figure~\ref{fig:scaling-edges} show that our
GPU curves steepen at high edge counts, while depthmapX maintains consistent
power-law scaling.
This reflects GPU resource saturation: at the city-scale run
(\num{12.1e9}~edges), batch streaming across PCIe and VRAM pressure cause
per-edge throughput to degrade.
Extrapolating both fits, the BFS-phase crossover---where depthmapX's
linear-memory exact BFS would match HyperBall---occurs around
$10^{12}$ edges, corresponding to roughly \num{50} million cells at
\SI{5}{\metre} spacing (a ${\sim}$\SI{35}{\kilo\metre}~$\times$~\SI{35}{\kilo\metre}
metropolitan area with unlimited visibility).
At that scale both tools would require years of compute, so the crossover
is theoretical rather than practical.
Moreover, the saturation point is hardware-dependent rather than
algorithmic: our benchmarks use a 2021-era laptop GPU with
\SI{16}{\giga\byte} VRAM and a PCIe~4.0~$\times$8 link---a modest
configuration by current standards.
Unified-memory architectures such as Apple Silicon, where GPU compute
shares the full system memory pool without PCIe streaming, would
eliminate the batch-transfer bottleneck entirely and push the
superlinear regime to substantially higher edge counts.

\paragraph{Future work.}
Warp-cooperative decoding of the LEB128 stream could further reduce
per-node overhead in the decode-union kernel.
Multi-GPU scaling would enable metropolitan-scale analysis at fine resolution.
A QGIS plugin wrapping the Rust core, in the spirit of the Space Syntax
Toolkit~\cite{gil2015qgis}, would make the tool accessible to
space-syntax practitioners without command-line expertise.
Multi-radius blending---weighting HyperBall results across several visibility
radii---is a natural extension that we defer to a follow-up study.

\section{Conclusion}
\label{sec:conclusion}

We presented a system that enables city-scale Visibility Graph Analysis by
combining delta-compressed CSR storage, HyperBall distance estimation, and
GPU acceleration, using the SparkSieve visibility algorithm ported from
depthmapX to ensure identical edge sets.
The most consequential property of HyperBall for VGA practice is that
computation time scales with topological depth: at radius-3---the standard
local measure in space syntax~\cite{hillier1984social,hillier1996space}---the
BFS phase achieves a $352{\times}$ speedup over depthmapX, whose BFS time is
invariant to depth setting.
This alignment between analytical best practice and computational efficiency
means that the analyses practitioners most commonly perform are also the
fastest to compute.
End-to-end, our tool achieves a $239{\times}$ speedup over depthmapX at
the largest matched configuration (\num{42705} cells, $p = 10$), and
scales to \num{236000} cells (\num{4.8e9} edges) in
\SI{137}{\second}.
At city scale, the full pipeline completes \num{2.7} million cells in
\SI{5.5}{\minute}.
Mean Depth accuracy is Pearson $r = 0.999$ (median relative error
\SI{1.7}{\percent}) at $p = 10$, and users can tune HLL precision to
trade accuracy for speed.
Together, these contributions open VGA to study areas and resolutions
previously infeasible with existing tools.

\bibliographystyle{unsrt}
\bibliography{references}

\end{document}